# Analysis and implementation of the Large Scale Video-on-Demand System

Soumen Kanrar

Vehere Interactive Pvt Ltd

Calcutta -53 India

Soumen.kanrar@veheretech.com

## ABSTRACT

Next Generation Network (NGN) provides multimedia services over broadband based networks, which supports high definition TV (HDTV), and DVD quality video-on-demand content. The video services are thus seen as merging mainly three areas such as computing, communication, and broadcasting. It has numerous advantages and more exploration for the large-scale deployment of video-on-demand system is still needed. This is due to its economic and design constraints. It's need significant initial investments for full service provision. This paper presents different estimation for the different topologies and it require efficient planning for a VOD system network. The methodology investigates the network bandwidth requirements of a VOD system based on centralized servers, and distributed local proxies. Network traffic models are developed to evaluate the VOD system's operational bandwidth requirements for these two network architectures. This paper present an efficient estimation of the of the bandwidth requirement for the different architectures.

## General Terms

Bandwidth optimization by using the user interactive traffic model.

## 1. INTRODUCTION
The proposed models provide a framework for studying the key parameters that have influence on the VOD network infrastructure. The research achievements are expected to be very useful for VOD system designers due to the essential roles of appropriate configuration and tuning of multiple VOD system parameters in achieving a cost-effective VOD system design. The network provides a new generation of multimedia services over high speed network and broadband technology that will support high definition television, and the demand for video content of DVD-quality. Thus, the video services for online deliberation will result from the merger of three industries: computing, communications, and broadcasting. Although it involves the many advantages, but there is a need for further research in order to achieve wide spread it. This is due to economic constraints and obstacles faced by design, as well as to the need for a large initial investment required to provide full service. This paper presents an overview and models for planning and selection of the network system based on demand video service. The proposed frame work provide a mechanism to determine the bandwidth requirements of the network, assuming the two approaches one for central network and the other for distributed. Models have been developed to determine the operational requirements of bandwidth of these two architects. These models also allow the possibility of studying the various factors affecting the infrastructure of the network. It is expected based on the research achievements to be very useful to designers of systems. The video on demand service given the need for proper preparation and the need for multiple network Setting transactions to ensure an effective network of appropriate economic cost.The TV delivery mechanism over NGN will support various services while maintaining its required levels of security, interactivity and reliability. It will allow users to access video services, such as entertainment movies, advertisements, electronic encyclopaedia, interactive games, and educational video documentaries from video servers on a broadband network. Potential new applications based on VOD systems include video information retrieval services, collaboration and conferencing systems, and distance learning [1].To meet and design objectives and to fulfill the requirements of VOD service, VOD systems require continuous data transfer over relatively long periods of time, media synchronization, very large storage, and special indexing and retrieval techniques adapted to multimedia data types. Therefore, a cost effective design for VOD system needs to evaluate a collection of various VOD system components [2]. In terms of the VOD transmission network, the system design must guarantee the required bandwidth for video traffic, and this bandwidth must support the VOD service quality needs and meet its packet loss polices. Video content delivery consumes large amounts of bandwidth in these networks due to its scalability, i.e. it must be able to support a large number of clients, thus imposing a heavy burden on the network and the system resources. A high definition (HD) stream, for instance, may require 10Mbps or more of bandwidth under MPEG-2 encoding. Therefore, any network link that handles many subscribers, each capable of demanding one or more VOD streams, must have enough bandwidth to meet the users' demands. In addition, system client must comply with the necessary buffer size and video request rate for the VOD delivery policy [3]. The Video on demand over broadband networks has been a prolific area of research [4-9]. The particular problems of provisioning VOD bandwidth, system deployment and architecture design have also received extensive attention in the literature [10-17].Reference [6] develops a performance evaluation tool for the system design and a user activity model to describe the utilization of network bandwidth and video server usage. An extensive survey on video-on-demand networks, their design approaches and future research challenges, is covered in [7]. The research in reference [11] builds a mathematical model for determining the TV bandwidth demand of multicast and surfing during commercial breaks. Various scalable VOD distribution architectures for broadband operators (based on a P2P streaming concept), together with schemes for VOD equipment allocation, are discussed in [12, 13] and [10, 14] respectively. This paper provides a planning methodology for analysing VOD architectures in order to determine their bandwidth requirements. Additionally, it examines all the key parameters that have influence on VOD communication networks. The remainder of this paper is organized as follows. In Section II, present the architectures and topologies that have been considered for VOD distribution networks, as well as the access mechanisms and traffic patterns required for retrieval of on-line video files. In Section III, present the planning methodology developed to estimate the VOD network bandwidth requirements. In Section IV, present a performance analysis and simulation parameters of the VOD network, discuss the various results obtained. In the last section, summarize with the acknowledgement and references.

## 2. COMPONENT ARCHITECTURE

This section describes the basic component required for the video on demand system.

### 2.1 Basic Components of the VOD System

A typical VOD broadband network consists of a number of remote client clusters that communicate their video requests via the network inbound links (client-to-server), and their video broadcasts via the network outbound links (server-to-clients). The components of the VOD network comprises the service control points, intelligent peripherals (such as multimedia storage servers, set-top boxes, cluster switches), and primary multimedia routers; these determine the system performance and communication costs. Stream based video delivery encodes all video, whether broadcast or video-on-demand, into data packets and transmits them to subscribers over the networks. VOD system architecture designs range from the simplest centralized system to complex distributed systems. The architecture design for VOD systems is based on the incorporation of such continuous media into a large array of extremely high-capacity storage devices, such as optical or magnetic disks, which are randomly accessible, with a short seek time, and are permanently on-line [4,8]. Video object delivery from the server to the client in general may be composed of multiple media streams, such as audio and video, whose retrieval must proceed so as to not only maintain continuity of playback of each of the constituent media streams, but also preserve the temporal relationships among them [5]. Various network architectures exist for producing VOD designs that minimize network costs and fulfill the service quality constraints. In this paper, main focus on two major widely used architectures: Centralized Network and Distributed Local Proxies. More on this is described in the following sections.

### 2.2 Centralized Network Architecture

In centralized architecture, all remote clusters communicate with the network's centralized primary servers through a broadband channel, which represent the backbone of the network; there are no local servers. All client requests are received by the primary server's router, which acts as a gateway where data decoding, de-multiplexing, regeneration, multiplexing, encoding, and carrier switching take place. The video primary servers then retransmit the video data information to the destination clients via the outbound links. The basic characteristic of the centralized video-on-demand system, as depicted in Figure 1, is that the multimedia information is always transported on demand from the central multimedia server to the subscribers through the network. If the server fails or becomes incapable of supporting existing connections, these connections will be blocked [3]. This solution suffers from very significant scalability problems, especially when scaled up for millions of potential users. Providing access to a large library of pre-encoded content using this approach requires enormous servers with enormous network connections [12].

### 2.3 Distributed Local Proxy VOD architecture

In distributed local proxy architecture, local proxy servers are installed at strategic locations in the network (closer to the clients). Remote clusters can communicate with the network's centralized primary servers as well as with its local proxy servers. Each local cluster server can support a number of customers connected to it through a cluster switch. The customers are connected to the central server's location through the cluster router, which acts as an interface between the client cluster and the broadband network. The main idea of distributed VOD local proxies (as shown in Figure. 2) is to distribute the centralized multimedia server functions within the network using the concept of local proxy storage. If the user cannot be served by the local proxy multimedia server for any reason, such as the blockage of the local multimedia server, or the multimedia information is not available in the local proxy server, then the request of the user will be transported to the centralized multimedia server. By locating the proxy server close to the user, it is expected that there will be significant reductions in the load on the system as a whole [7-8]. Another advantage of the distributed local proxy video-on-demand system is that it can be expanded in a horizontal manner for system scalability and evolution. It can start from an initial two-level system (with a centralized multimedia server and one local video server) to a system with as many local servers as needed. Compared with the centralized multimedia server system, the distributed system may utilize a lower than average network bandwidth and have higher system reliability, but at the expense of needing a significant amount of local storage systems.

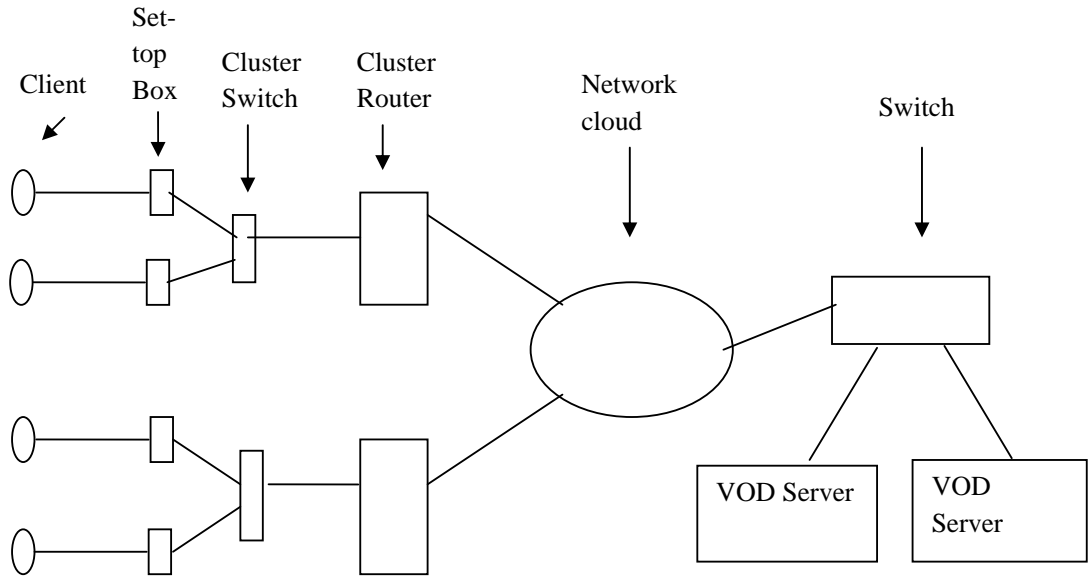

Figure 1: Video-on-demand Network (Centralized)

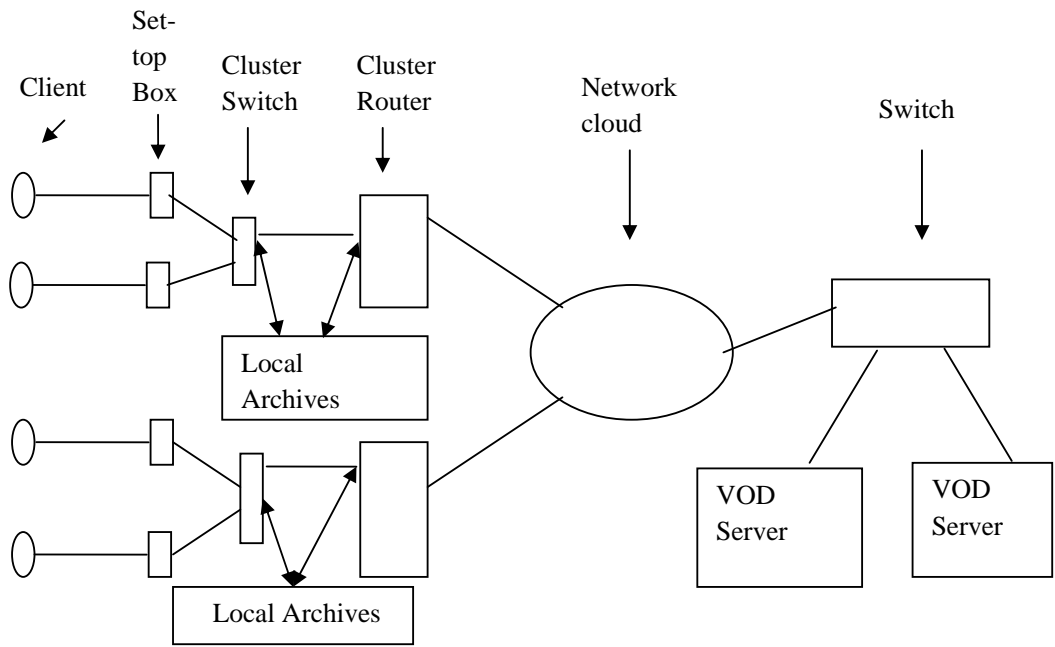

Figure 2: Video-on-demand Network (Distributed)

## 3. VOD Traffic Models

Once connected, we assume the user is in one of two states, the normal or the interactive state. He starts in the normal state, i.e. the video is being played at the normal speed. He stays in this state for a period of time, which is exponentially distributed with parameter $\alpha$, and then the user issues an interactive operation, such as stop, speed up, etc. He stays in this interactive state for period of time, which is also exponentially distributed with parameter $\beta$. Then he goes back to the normal state, from where he may again go to the interactive state. This may be repeated many times until he disconnects. Different types of interactive operations will affect the system in different ways, and these include various common VOD watching operations such as: Play/Resume, Stop, Pause, Jump Forward, Jump Backward, etc. [6].The study for engineering VOD links will assume stationary (steady state) traffic of user activity. In the steady state, normal requests for a video movie are serviced through a multicasting mechanism where viewers who watch the same broadcast movie share one stream of broadcasting, which lasts for an average period of ($t_n$) minutes (equivalent to the normal request exponential with parameter $\alpha$) For movie surfing, the steady state assumption is that the surfers' interactive requests are serviced by the network through sending a uncast (one per viewer) short-lived stream, each for an average period of ($t_i$) minutes (equivalent to the interactive request exponential distribution with parameter $\beta$) Due to their uncast nature, each interactive request superimposes a significant additional demand on top of the steady state normal demand, and therefore the bandwidth capacity planning and provisioning must include surfing effects. Throughout the study, the average client view activity is assumed, as shown in Table 1, but represents typical empirical data [17]. Our study will cover two types of service available to VOD system users, namely: high definition (HD) and standard definition (SD) VOD movies.

**Table 1**: **VOD service features assumed in traffic analysis**

| VOD Service Feature | Value |
|---|---|
| Peak viewing period | Sat. nights (3to 4 hrs/night) |
| Peak busy per subscriber | 7hrs |
| Normal (steady state) request attempts per movie per period | 1 to 5 times |
| Holding time of interactive request | 1 to 10 seconds |
| Interactive-request attempts per movie per period | 1 to 4 times |
| Bandwidth per port | Variable; 3Mbps-20Mbps |
| SD movie bandwidth | 3Mbps |

### 3.1 Mathematical Model of the Client Request

The distribution of VOD movies request generally follows a Zipf-like distribution, where the relative probability of a request for **i** ( the most popular page) is proportional to $1/i^\alpha$, with $0 < \alpha < 1$ and typically taking on some value less than unity. The request distribution rarely follows the strict Zipf law (for which $\alpha$ =1) [18].For Zipf-like distribution, the cumulative probability that one of the K popular movies is accessed (i.e. the probability of a popular movies request) is given asymptotically by

$$\Psi(k) = \sum_{i=1}^{K} \frac{\delta}{i^\alpha} \approx \delta \frac{k^{1-\alpha}}{(1-\alpha)} \quad \text{.......(1)}$$

$$\text{and } \delta \approx \frac{(1-\alpha)}{N^{(1-\alpha)}}$$

where N is the total number of movies in the system.

$\Psi(k)$ can be approximated as:

$$\Psi(k) \approx (k/N)^{(1-\alpha)} \quad \text{.......(2)}$$

Because $k/N \prec 1$ for all meaningful $k$, and a larger $\alpha$ increases $\psi(k)$, most requests are concentrated on a few popular movies [95]. Based on this, we can estimate the probability of a request for an unpopular movie, for a VOD system with $N$ total movies and $k$ popular ones, as:

$$P_{unp} = 1 - (k/N)^{1-\alpha} \quad (3)$$

The analytical methods for provisioning links in this study assume steady state busy hour traffic for movie retrieval. In the steady state, multicasting is used to reduce VOD traffic volumes. The network needs to deliver only one video stream (one video server port) for a group of viewers (multicast group) watching the same video or broadcast program segment. The steady state demand is therefore the total bandwidth of all video streams (or server ports) in use. There are two types of request in the VOD network; the first one is the request for initializing or starting the video movie (labelled normal request in the study). The other type is the request for interactive service (e.g. stop/pause, jump forward, fast reverse, etc) to be performed on the viewed movie (labelled interactive request in the study). Since each of these requests is independent from each other, and the arrival requests come from large numbers of client set-up terminals, the arrival process of normal requests, as well as of interactive requests, to each video server can be modelled as a Poisson process with average rates $\lambda_n$ and $\lambda_i$ respectively. With this assumption, the distribution of the sum of $k$ of independent identically distributed random variables, representing the request inter arrival times, is then the Erlang distribution.

### 3.2 Analytic Model for Centralized Architecture

In this section we present the traffic models used to determine the VOD system bandwidth required for a 'no blocking' service. The model is based on the Erlang-B formula with different values of blocking probability. Our aim is to estimate the number of server ports supported by the down broadband link channel (from video server to client set-up box). Using the estimated number of server ports, the VOD system bandwidth required is then simply determined by multiplying the movie rate (according to the movie being HD or SD) by the number of server ports, determined from the Erlang -B formula. The total VOD system demand equals the sum of the server ports in use times the movie bandwidth per port stream. We proceed as follows:

Let $Mc$ = Total network traffic in Erlang for a centralized system.

$$M_c = \frac{(x).(h).(p).(\lambda_n).(t_n)}{(Z).T} + \frac{(x).(h).(p).(\lambda_i).(t_i)}{T} \quad (4)$$

Where $x$ = Number of VOD system cluster areas,

$Z$ = Multicast factor (i.e. number of viewers who request the same multimedia movie within a short period of time, thus it can be served from the same server port),

$h$ = Number of houses in VOD system cluster service area.,

$p$ = Penetration of service in a VOD system cluster area,

$\lambda_n$ = Average number of normal request attempts per movie per period per household,

$\lambda_i$ = Average number of interactive request attempts per movie per period per household,

$t_n$ = Holding time of a normal request for a movie in minutes,

$t_i$ = Holding time of an interactive request in minutes,

$T$ = Peak busy period in minutes.

Now, the number of server ports supported by the VOD network ($Sc$) can then be found using the Erlang-B formula, from the total calculated network traffic ($Mc$) with a given blocking probability $P_B$, where

$$P_B = \frac{\frac{(M_C)^{S_C}}{(S_C)!}}{\sum_{N=0}^{S_C} \frac{(M_C)^N}{N!}} \quad (5)$$

And the requires bandwidth is then

$$W_c = S_c * r \quad (6)$$

Where $r$ is the movie rate (e.g. 3Mbps for SD movie),

and the allocated bandwidth per household ($W_{ch}$) is given by

$$W_{ch} = \frac{(r*S_c)}{(x*h)} \quad (7)$$

### 3.3 Distributed VOD Architecture

In this section we conduct a traffic analysis to determine the number of servers needed for a distributed local proxy IP/VOD architecture. The assumption is that the system will store the unpopular materials on a high capacity optical disk, while the popular materials are stored as on-line mass storage, in both the local and the centralized servers.

To simplify the analysis, we assume that most of the cluster-area traffic through the broadband link is for non-popular requests. Therefore, there is no blocked popular request traffic at the local server. This assumption is justified as it is expected that local servers will contain all popular movies. We proceed as follows:

$MC_L$ = Total traffic supported by broadband network from all cluster areas in the distributed local system,

$M_L$ = Local traffic at the local cluster area,

$S_L$ = Number of server ports supported by the

Cluster area VOD local servers,

$S_{CL}$ = Number of server ports supported by the centralized primary servers,

$P_{un}$ = Probability of unpopular request, which is given by Equation 3 as

$$P_{un} = 1 - (k/N)^{1-\alpha}$$

Now $M_L$ can be found from:

$$M_L = \left[\frac{h.p.\lambda_n.t_n}{Z.T} + \frac{h.p.\lambda_i.t_i}{T}\right].(1 - p_{un}) \quad (8)$$

With, $t_n, t_i, \lambda_n, \lambda_i, T, Z, x$, and $p$ as defined before in Equation 4.

$S_L$ can be found from $M_L$ using the Erlang-B formula for a given blockage probability $P_B$ as:

$$P_B = \frac{\frac{(M_L)^{S_L}}{(S_L)!}}{\sum_{N=0}^{S_L}\frac{(M_L)^N}{N!}} \quad (9)$$

Now $M_{CL}$ is calculated as follows:

$$M_{CL} = [\frac{h.p.\lambda_n.t_n}{Z.T} + \frac{h.p.\lambda_i.t_i}{T}].P_{un} \quad (10)$$

Similarly we can find $S_{CL}$ from $M_{CL}$ using the Erlang-B formula. The local cluster-area bandwidth $W_{LL}$ and the bandwidth to central servers $W_{LC}$ can be found by multiplying the corresponding number of ports with the movie bandwidth. Finally the total required server ports per cluster area are then the sum of $S_{CL}$ and $S_L$ and the total bandwidth per area is given by:

$$TW_{LC} = r*(S_{CL} + S_L) = W_L + W_{LC} \quad (11)$$

Then, the allocated bandwidth per household $W_{ch}$ is given by:

$$W_{ch} = r*(S_{CL} + S_L)/h \quad (12)$$

The overall system bandwidth is

$$TW = x*TW_{LC} \quad (13)$$

*3.4 VOD Inbound Channel Requirements*

In this section we study the requirements of the network inbound channel (from cluster-area clients to VOD servers) that support the proposed VOD system. This is needed as client links may be implemented using asymmetrical connection technologies such as ADSL or VSAT. Such links work under the assumption that the clients will download more information than they send, and for the VOD clients, the analysis includes an evaluation of the requirements in terms of the number of cluster areas that can be supported as a function of request message length and rate.

Let the inbound channel bandwidth = $W_u$, the number of bits per normal request = $L_n$, the number of bits per interactive request = $L_i$, interactive request rate per house = $\lambda_i$, normal request rate per house = $\lambda_n$, T be the busy peak period, and $P_{un}$ = probability of unpopular request.

We calculate $W_{OC}$ for the VOD (both centralized and local proxy) systems as follows:

$$W_{oc} = \frac{Pxh(\lambda_n L_n + \lambda_i L_i)}{T} \quad (12)$$

## 4. Results and Performance Analysis

In this section I conduct the traffic analysis to determine the VOD system's inbound and outbound channel bandwidths required for a 'no blocking' service. The traffic analysis is carried out using the developed traffic models with different values of blocking probability. The VOD system is analyzed for various video movie resolutions including standard definition VOD (video and audio) and high definition VOD with an average rate per port of 3Mb/s and 8Mb/s respectively.

*4.1 Simulation Parameters for the Centralized System*

The effects of some of the centralized system parameters (such as the number of clusters, interactive session like movie holding time, Pause, fast, fast forward and the average number of requests arriving to the system during the peak period shown in table 1.

**Table 2: Parameters related to real world scenario**

| No. of Cluster | 40 | 50 | 60 | 70 | 80 | 90 |
|---|---|---|---|---|---|---|
| No. Of Ports | 7000 | 6530 | 6100 | 5300 | 4800 | 4200 |
| Traffic Intensity | 8700 | 8100 | 7430 | 6700 | 6100 | 5300 |
| Bandwidth in Gbps | 30 | 27 | 24 | 22 | 19 | 15 |
| Bandwidth /house-in Kbps | 173.8 | 163.9 | 145.3 | 132.7 | 127.3 | 112.5 |
| $W_C$ Gbps | 52.4 | 46.7 | 43.5 | 41.9 | 39.4 | 38.1 |
| $W_{ch}$ KBPS | 310.2 | 305.9 | 301.2 | 297.5 | 288.5 | 280.1 |

The table 2 represents the parameters related to the real world scenario of the centralized video on demand system. The other parameters are assign like maximum number of clients in the cluster i.e. X = 250 , h = 600, traffic rate $\lambda_n$ = 2.5, $\lambda_i$ =4, 6,8 and the interactive session for $t_n$ = 120 seconds , assuming $t_i$ =6, admissible blocking probability =.05. Simulation runs for 5 minutes.

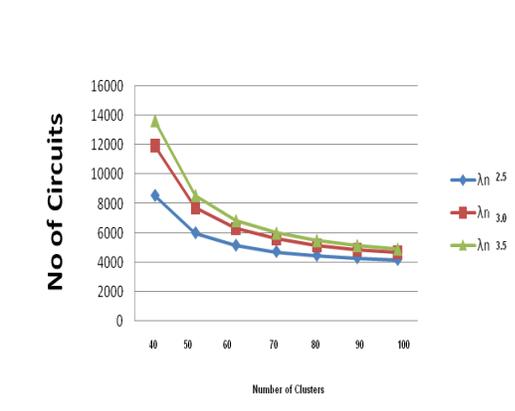

Figure 3

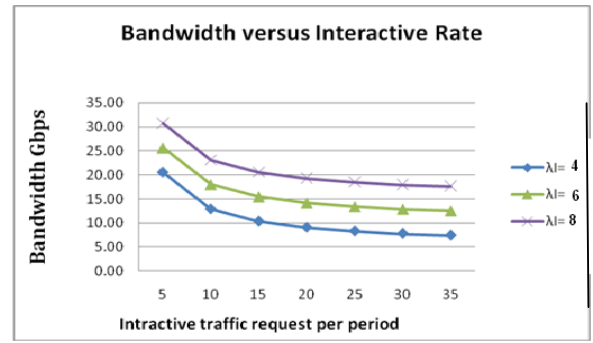

Figure 4

It is clear from the figure 3 that the VOD systems with the help of clustering significantly reduce the required bandwidth. For example if there is no cluster i.e. number of cluster is 1 means, bandwidth be consumed as unicast streaming as peer to peer network. Also, high video movie resolution required bandwidth significantly. In Figures 4 and 5, represents the relationship between the required channel link required to serve the clustered customers. The figures 4 shown the required bandwidth comes down for any types of the interactive session pause/skip/hold/fast-forwards, when the number of cluster is increases. In the simulation, we have considered the total number of clients in the population is fixed. The sum of the clients belongs to each clusters is equal to the total client population size. The bandwidth of the system is also depends on the size of the client population for a given resolution. The blocking probability is used to reduce the load in side server. If the client request forwarded to the server buffer, the client request put into the buffer because already the resources occupied by the other clients. In that case the client is in long delay and unnecessary burden on the storage system, other important area is it will consume the important bandwidth. The blockage probability has great importance to reduce load inside the server i.e. used to load balancing. The blockage probability also used to enhance the overall performance by the full use of the bandwidth. Figure 5, presents the relationship between the client densities inside the cluster the required channel bandwidth for the interactive traffic request rate. The increment of the interactive traffic, (i.e. increment of $\lambda_i$) due to the random surfing for movies).

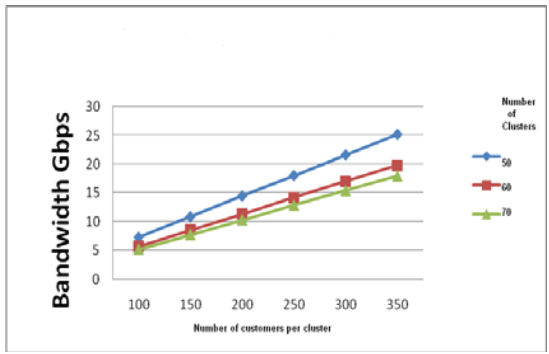

Figure 5

**Table 3: Parameters for the Distributed VOD system**

| No of Cluster | 40 | 50 | 60 | 70 | 80 | 90 | 100 |
|---|---|---|---|---|---|---|---|
| $S_L$ | 47 | 43 | 39 | 36 | 31 | 27 | 23 |
| $M_L$ | 33 | 31 | 27 | 23 | 21 | 19 | 17 |
| $M_{LC}$ | 9 | 7 | 5 | 4 | 3 | 2 | 1 |
| $S_{CL}$ | 22 | 29 | 17 | 15 | 11 | 7 | 4 |
| $W_{LL}$ | 201 | 186 | 172 | 135 | 116 | 108 | 101 |
| $W_{LC}$ | 87 | 81 | 72 | 66 | 52 | 44 | 41 |
| $W_{Ch}$ | .87 | .76 | .71 | .68 | .62 | .51 | .49 |
| $TW_{LC}$ | 203 | 192 | 181 | 172 | 161 | 150 | 141 |

The other parameters are assign like maximum number of clients in the cluster i.e. X = 250, h = 600, $\lambda_n$ = 2.5, $\lambda_i$ =4, 6,8 and the interactive session is $t_n$ = 120 seconds , $t_i$ =6, blocking probability =.05. Simulation runs for 5 minutes.

The system parameters (such as the number of clusters, interactive session during the peak period on the distributed local proxy system and the channel bandwidth has shown in Table 3. In Figures 6 and 7, presents the relationship between the channel bandwidth and the system normal video request rate and the growing cluster size. Figure 6 presents bandwidth requirement with the client request for local VOD architecture. The Client request generally follows Gaussian type statistical distribution. In general by increasing the size of the normal request rate will require more servers to satisfy the system blocking probability that presents in the figure 7. Figure 8 represent the system bandwidths for the local proxy architecture as a function of interactive session. Here the Clustering shown the greater importance in real scenario according to figure 8.

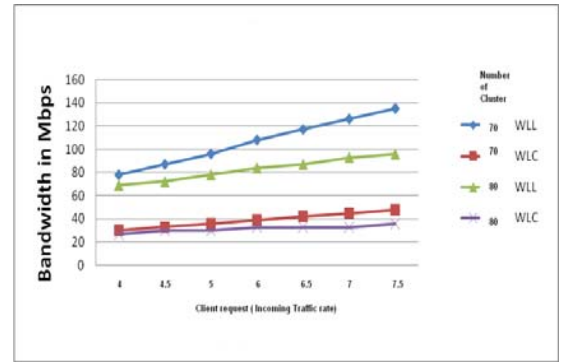

Figure 6

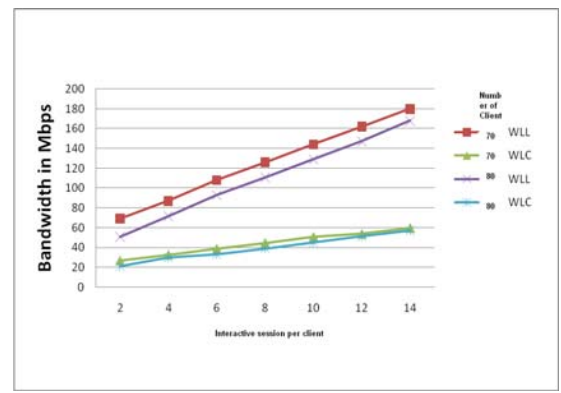

Figure 7

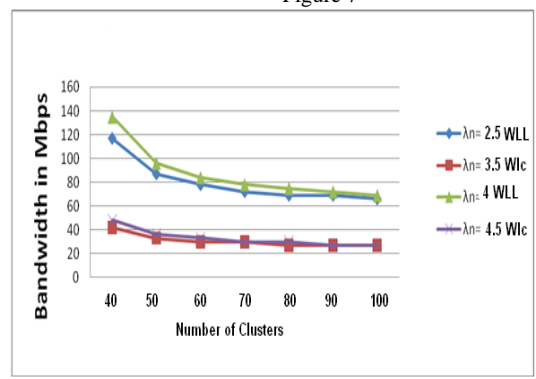

Figure 8

## 5. FURTHER IMPROVEMENTS

The dynamic spreading cluster and the dynamic request filtering will be added to the extended work. For the dynamic spreading cluster spectrum and the efficiently request handling, new algorithms required. The optimum strategy is also be reviews the help of the traffic scenario.

## 5. ACKNOWLEDGMENTS

My thanks to the unknown experts who have contributed towards development of the Manuscript.

**Author**

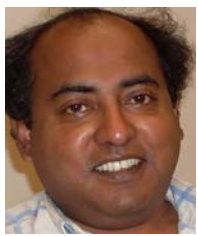

**Soumen Kanrar** received the M.Tech. degree in computer science from Indian Institute of Technology Kharagpur India in 2000. Advanced Computer Programming RCC Calcutta India 1998. and MS degree in Applied Mathematics from Jadavpur University India in 1996. BS degree from Calcutta University India. Currently he is working as researcher at Vehere Interactive Calcutta India. Previously he had worked at King Saud University, Riyadh. Formally attached with the University Technology Malaysia. He is the member of IEEE.